\documentclass[12]{amsart}
\usepackage{graphicx}
\usepackage{amssymb}
\usepackage{amsaddr}
\usepackage[usenames]{color}
\numberwithin{equation}{section}
\usepackage[hypertex,colorlinks]{hyperref} 
\definecolor{darkblue}{rgb}{0.0,0.0,0.3}
\hypersetup{colorlinks,breaklinks, linkcolor=darkblue,urlcolor=darkblue,      anchorcolor=darkblue,citecolor=darkblue}
\title[]{Magnetic properties of a Fermi gas in a noncommutative phase space}%
\author{S.\ Franchino Vi\~{n}as\quad and \quad F.\ Vega}%
\address{Departamento de F\'isica, Facultad de Ciencias Exactas,\\ Universidad Nacional de La Plata, C.C. 67 (1900), La Plata, Argentina}%
\email{safranchino@gmail.com\quad-\quad federicogaspar@gmail.com }

\begin{document}
\maketitle
\thispagestyle{empty}
\vspace{2cm}
\begin{abstract}
Motivated by the precision attained by SQUID devices in measuring magnetic fields, we study in this article the thermodynamic behaviour  of a fermion gas in two and three dimen\-sional spatial space with noncommutative coordinates and momenta. An explicit expression, both for Landau's diamagnetism and Pauli's paramagnetism, is obtained for the magnetization and magnetic susceptibility of the gas in two and three spatial dimensions. These results show that an upper bound for the noncommutative parameter $\theta\lesssim (10 \,\text{Gev})^{-2}$ could be obtained.
\end{abstract}


\newpage
\pagenumbering{arabic}
\section{Introduction}

Currently noncommutative (NC) theories are of manifest interest in a wide range of areas such as geometry \cite{Connes1}, condensed matter \cite{Dayi:2001nv,Bellissard,Falomir:2011tf,Cabra:2005fj} and quantum gravity (both from the theoretical \cite{Seiberg:1999vs,Witten:1985cc} and phenomenological \cite{Hinchliffe:2002km,AmelinoCamelia:2008qg,Falomir:2013vaa} point of view).  In this framework, one of the most important motivations for studying NC spaces comes from string theory. In effect, in the works of  Connes, Douglas \& Schwarz \cite{Connes:1997cr}, Seiberg \& Witten \cite{Seiberg:1999vs} and Douglas \& Hull \cite{Douglas:1997fm} it was shown that several low energy limits of string theory and $M$-theory correspond to effective field theories in NC spaces. Given the technical difficulties encountered in the ambitious unification programme of gravity and quantum field theory \cite{Szabo:2001kg, Doplicher:1994tu} and with the aim of constraining and detecting beyond-the-standard-model effects, NC
quantum field
theories and specially NC quantum mechanics (NCQM)  have been subjects of intense research in the last ten years.


Of course, singular effects of spatial noncommutativity, such as the generalized Heisenberg uncertainty relation, are expected to become relevant only near Planck's scale. 
Nevertheless, it is also conjectured that some low energy relics of such effects may be verified by current experiments. Needless to say, if it is possible to find evidence of noncommutative physics it will be probably found in the most precise experiments. As a consequence, the exploration of some  quantum mechanics solvable models in NC space is of crucial phenomenological importance.

Several works have been devoted to this topic -- in particular to establish bounds for parameters of noncommutativity. Indeed, strong bounds of the order of Planck's energy can be obtained from astrophysics phenomena, such as gamma-ray bursts \cite{Vasileiou:2015wja,Borowiec:2009ty}. Some weaker bounds ($E\sim 10^2 \text{TeV}$) were computed  in \cite{Horvat:2009cm} analysing the primordial nucleosynthesis and in \cite{Carroll:2001ws}, where the bound comes from Lorentz invariance violation. Values slightly lower than these were found by considering NC radiative corrections to the Lamb shift \cite{Chaichian:2000si}, the Aharonov-Bohm NC problem \cite{Falomir:2002ih} and NC Bose-Einstein condensates \cite{Gamboa:2009zs}.


\medskip

On the other side, one of the most precise experimental devices are SQUIDs -- magnetometers made of Josephson junctions capable of measuring fields as small as $1\,\text{aT}$, thanks to the smallness of the superconducting magnetic flux quantum $\Phi_0=\frac{h}{2e}$. Following our line of reasoning, we conclude that an analysis of magnetic fields in noncommutative geometries should be performed. Therefore, we will analyse in the present article the thermodynamics of a NC fermion gas in the presence of a constant external magnetic field in $d=2$ and $3$ spatial dimensions. Our final goal is to estimate the contributions to the magnetization and susceptibility of the gas coming from the position and momenta NC parameters which is expected to be of the order of Planck's length, by considering spin $0$ and $1/2$ fermions. Afterwards, an upper bound for the NC parameter $\theta$ is obtained in terms of the SQUID resolution.

Some systems including magnetic fields in NCQM have been previously studied in \cite{Delduc:2007av} and in conjunction with an harmonic term in \cite{Bellucci:2003xd,Nouicer:2007hi,Jellal:2001nu}. As far as we know some thermodynamical
quantities were only computed in \cite{Dayi:2001nv} and \cite{Jellal:2001nu} -- however we consider in addition the low temperature regime, Pauli's paramagnetism and a noncommutative parameter for the momenta operators.
\medskip

This paper is organized as follows: we present in Section \ref{review} a brief review regarding the spectra of a Schr\"odinger particle in a NC space subject to a constant magnetic field\cite{Falomir:2015una}. Next, we compute the grand canonical partition function for a spinless NC fermion gas in the low and high temperature regimes. In particular, explicit expressions for the magnetization and magnetic susceptibility are obtained. In section \ref{section.pauli} we explore the effects of Pauli paramagnetism by adding an interaction term between the spin $1/2$ fermions and the external field, and the new contributions for the magnetization and the susceptibility are computed. Finally we discuss on the possibility of measuring this NC corrections. In appendix \ref{apendice-fugacidad} and \ref{apendice-abel-plana} we present some technical details regarding the computation of the fugacity and the grand canonical partition function in the low temperature regime.


\section{Small review of the Landau Problem in the NC plane}\label{review}

Consider a Schr\"odinger particle with charge $e$ and mass $m_e$, living on a 2D plane and  subject to a constant magnetic field $B$ chosen to be perpendicular to the plane. The Hamiltonian $\mathcal{H}$ of this system, in the symmetric gauge, can be expressed as
\begin{equation}\label{Landau-1}
      \begin{array}{c} \displaystyle
         \mathcal{H}:=\frac{1}{2\,m_e}\left( \hat{\pi}_i+\frac{e B}{2}\, \epsilon_{ij} \hat{q}_j\right)^2,
      \end{array}
\end{equation}
where $\epsilon_{ij}$ is the Levi-Civita symbol and for definiteness we take $e B > 0$. This is usually called the Landau problem. Its 3D version is not harder to solve: if we had chosen the particle to live in a 3D space, the Hamiltonian could have been divided into this 2D Hamiltonian on a plane plus a free particle in the perpendicular direction.

The extension of the 2D problem to the NC plane with nonvanishing commutators between both coordinates and momenta operators, $\hat{q}_{i}$ and $\hat{\pi}_{i}$, is characterized by the following commutation relations\footnote{From now on we will set $\hbar=1$.}:
\begin{equation}\label{1}
      \left[\hat{q}_i , \hat{q}_j\right] = \imath \theta\, \epsilon_{ij} \,, \quad
      \left[\hat{\pi}_i , \hat{\pi}_j\right] = \imath \kappa \, \epsilon_{ij} \,,\quad
      \left[\hat{q}_i , \hat{\pi}_j\right] = \imath \,  \delta_{ij} \,.
\end{equation}
In this expression $\theta$ and $\kappa$ are the (real) noncommutativity parameters and without loss of generality, we can take $\theta\geq0$. For simplicity we will also consider $\kappa>0$.

For the analogous NC 3D problem a remark is now in order. If the NC-matrix $\Theta_{ij}=[q_i,q_j]$ is made of constants it can be shown that the space reduces to a \emph{NC-plane} times the perpendicular commuting direction. The same may be said about the noncommutativity in the momenta. However, it may be the case that these resulting noncommuting coordinates and momenta are not canonically conjugated -- if they are canonically conjugated, the problem reduces to a problem in the \emph{NC-plane} plus a free particle in the perpendicular direction. We will always consider this to be the case and the field to be also perpendicular to the \emph{NC-plane}.

An interesting way to obtain the spectrum of nonrelativistic rotationally invariant Hamiltonians such as \eqref{Landau-1} is by finding conserved quantities. Of course, the generator of rotations on the \emph{NC-plane} is conserved and given by \cite{Nair:2000ii}
\begin{equation}\label{gen-rot}
    \hat L:=\frac{1}{\left(1-\theta\kappa\right)}\left\{\left( \hat q_1 \hat\pi_2 -\hat q_2 \hat\pi_1\right)+
   {{  \frac{\theta}{2 } \left({\hat\pi_1}^2+{\hat\pi_2}^2\right)}+{ \frac{\kappa}{2}
   \left({\hat q_1}^2+{\hat q_2}^2\right)}}\right\}\,,
\end{equation}
for $\kappa \theta\neq 1$.
For the critical value $\kappa_c=\theta^{-1}$ (in which case $\hat{L}$ has no sense), \eqref{1} can be satisfied by a single pair of dynamical variables. Therefore, for this particular value there occurs an effective \emph{dimensional reduction}.

As it was demonstrated in \cite{Falomir:2015una}, every Hamiltonian with a central potential in the \emph{NC-plane} has another conserved quantity corresponding to the quadratic Casimir invariant, whose eigenvectors transforms according to irreducible representations of $SL(2,\mathbb{R})\otimes SO(2)$ or ${SU(2)\otimes SO(2)}$ according to whether $\kappa$ is lower or greater than a critical value $\kappa_c=\theta^{-1}$. Under  these conditions, the eigenvalue problem for the Hamiltonian depends on the region we are considering and therefore we shall study these two regions separately.
Now on, we will refer to this regions as I or II respectively.

\subsection{Region I: $\kappa<\kappa_c$ case.}

Here, the Hamiltonian shall be written (see \cite{Falomir:2015una}) in a space of the (infinite dimensional) unitary irreducible representations (\emph{irreps}) $\langle k,l \rangle$  of ${SL(2,\mathbb{R})\otimes SO(2)}$. Among these existent \emph{irreps}, those compatible with our problem involve only discrete classes of $SL(2,\mathbb{R})$ and live in spaces generated by the vectors $\lvert k,k+n, l \rangle$. In this notation, $l$  is a label for the angular momentum, $k$ a positive integer or half-integer and $n\in\mathbb{N}_0$. Then, it is straightforward to see that the spectrum of the Hamiltonian in this \emph{NC-plane} is equivalent to its commutative counterpart with an effective magnetic field\footnote{In the case $\kappa=0$ this effective field equals the one computed using Seiberg-Witten map for NC gauge field theories\cite{Scholtz:2005yv}.}
\begin{align}\label{effective-magnetic-field}
\mathcal{B}:=B+\left(  B^2 e^2 \theta +4 \kappa  \right)/(4e).
\end{align}
Indeed, the eigenvalues are given by
\begin{equation}\label{Landau-5-1}
    E_n^{(|l|,s)}=
     \omega_c\left(n+\frac{|l|}{2} (1-s)  +\frac{1}{2}\right)\,,
\end{equation}
where  ${\omega}_c=\frac{ |e \mathcal{B}|}{m_e}$ is the effective cyclotron frequency and $s=sgn(l\,.\mathcal{B})=\pm1$.

However, as it will be shown in section \ref{sec.density-states}, the density of states is not the same compared to the the commutative one. Notice also that $\kappa$ plays a similar role to a magnetic field -- indeed if $B=0$ for $\kappa\neq0$, then $\mathcal{ B}\neq0$. This has already been pointed out in \cite{Nair:2000ii}.

Another fact worth mentioning is the degeneracy of the levels. From one hand, if $s=1$ the eigenvalues are independent of $|l|$ and this accounts for a countable infinite degeneracy. From the other, if $s=-1$ we may redefine $n'=n+|l|$ and therefore the states with negative $l$  contribute to the characteristic subspace of the Hamiltonian with energy $E_n^{(|l|,+1)}$ with a finite number $n'$ of additional linearly independent eigenvectors.

\subsection{Region II: $\mathbf{\kappa>\kappa_c}$ case.}
Here, the eigenvalue problem for the Hamiltonian operator is reduced to a finite-dimensional one in the $\langle j,l\rangle$ unitary irreducible representation of $SU(2)\otimes SO(2)$ (see \cite{Falomir:2015una}). The states are also labeled with three indices $\lvert j, m, l \rangle$, where $2j\in \mathbb{N}_0$, $-j\leq m\leq j$ and the angular momentum is constrained to be $l = -(2j + 1)$. Furthermore, the eigenvalues of the Hamiltonian are given by
\begin{equation}\label{Landau-5}
    E_l^{(m)}=
     \omega_c\left(m + j + \frac{1}{2}\right)\,.
\end{equation}

Notice that these eigenvalues depends only on the nonnegative integer $n_a=j+m$. Therefore, given $n_a$, each irreducible representation with $j \geq n_a/2$ contains a state with energy $\omega_c\left(n_a +\frac{1}{2}\right)$ -- these states have therefore a countable infinite degeneracy and will also be computed in section \ref{sec.density-states}. 

Once again, comparing this spectrum with the Landau problem on the usual commutative plane, one can see that noncommutativity gives raise to the effective external magnetic field $\mathcal{B}$ and a corresponding density of states. Recall that in this region only negative integer values $l\leq -1$ are available for the angular momenta.

\subsection{Density of states}\label{sec.density-states}
Since we are interested in analysing the thermodynamic behaviour of a fermionic gas subject to a constant magnetic field we will compute the logarithm of the grand-canonical partition function $\mathcal{Z}$, defined as a sum over all the available states $n$
\begin{align}\label{funcion_particion}
\log \mathcal{Z} := \sum_{n} \log\left(1+ze^{-\beta E_n}\right),
\end{align}
where $E_n$ is the energy of the $n$-th state and $z=e^{\beta \mu_c}$ is the fugacity, being $\mu_c$ the chemical potential. A problem arises when one tries to compute the sum when the gas is contained in an infinite-volume container because of the energy levels degeneracy. A way to surpass this obstacle is to consider a gas with $\bar{N}$ mean particles contained in a finite volume $V$. Then one can divide both sides of eq. \eqref{funcion_particion} by $V$ -- as a consequence a natural density of states arises in the thermodynamical limit, i.e. in the $V\rightarrow\infty$ limit as the mean density $\bar{n}:=\frac{\bar{N}}{V}$ of particles remains finite. Since the way to derive the density of states $\rho$ in both regions I and II is analogue, we shall perform the proof in region II, viz. the region with $\kappa>\kappa_c$.

As a first step, we begin by counting the number of eigenstates contained within a circle of radius $R^2$ and centered in the origin, having in mind that the thermodynamical limit will be achieved by taking the $R\rightarrow\infty$ limit. In order to do so, we can compute the expectation value of the square $\hat{X}^2$ of the position operator -- for a given state $\vert j,n_a-j,-(2j+1)\rangle$ it reads
\begin{align}
 \langle \hat{X}^2\rangle_{j,n_a}^{II} = A^{-1} j+C\,n_a+\theta.
\end{align}
In this equation we have introduced the coefficients
\begin{equation}
A:=\frac{e\,\mathcal{B}}{4\,(\theta\kappa-1)}\,,\quad  C:=2\theta-A^{-1}\,.
\end{equation}

This result allows us to determine the number of states concentrated in a circle of radius $R$ by imposing the restriction $\langle \hat{X}^2\rangle_{j,n_a}\leq R^2$ on the states.
In this case we obtain for the grand canonical partition function the result
\begin{align}
\frac{1}{V}\log \mathcal{Z} = \frac{1}{\pi\,R^2}\sideset{}{'}\sum_{j,n_a} \log\left(1+ze^{-\beta\omega_c(n_a+\frac{1}{2})}\right),
\end{align}
where the prime on the sum means that the indices $j$ and $n_a$ are subject to the restrictions
\begin{equation}
\left\lbrace\begin{array}{rcl}
2j&\geq & n_a\\
R^ 2 &\geq &A^{-1} j+C\,n_a+\theta
\end{array}\right.\,.
\end{equation}

If we first perform the sum over the $j$ index, we arrive at the result
\begin{align}
 \frac{1}{V}\log \mathcal{Z} = \frac{1}{\pi\,R^2}\sum_{n_a=0}^{n_0} 2\left[ \lfloor A \left(R^ 2 -C\,n_a+\theta\right)\rfloor-\frac{n_a}{2}  \right]\;\log\left(1+ze^{-\beta\omega_c(n_a+\frac{1}{2})}\right).
\end{align}
In this expression we have made use of the floor function $\lfloor x \rfloor$ --which gives the biggest integer number smaller than or equal to $x$-- and we have defined $n_0:= \frac{2A(R^2-\theta)}{2CA+1}$. In the $R\rightarrow \infty$ limit we are left with the expression
\begin{align}
-\beta\phi_0:=\lim_{V\rightarrow\infty} \frac{1}{V}\log \mathcal{Z} = \sum_{n_a=0}^{\infty} \frac{2\,A}{\pi} \;\log\left(1+ze^{-\beta\omega_c(n_a+\frac{1}{2})}\right)
\end{align}
for the grand potential density $\phi_0$, from which one may instantly recognize the density of states
\begin{align}\label{density-II}
\rho_{II}=\frac{e\,\mathcal{B}}{2\,\pi\,(\theta\kappa-1)}.
\end{align}

As we have already stated, the density of states in the region I could be analogously computed and in the thermodynamical limit only states with $l=(2k-1)$ give a non vanishing contribution. The resulting expressions for the expectation value of the square of the position operator --for a given state $\vert j,n_a-j,-(2j+1)\rangle$-- and the density of states can be found to be
\begin{align}
  \begin{split}
\langle \hat{X}^2\rangle_{k,n_a}^{I} &= \frac{4\,(1-\theta\kappa)}{e\,\mathcal{B}} k+2\frac{e\,\theta\mathcal{B}+2(1-\theta\kappa)}{e\mathcal{B}}\,n_a+\theta,\\
\rho_{I}&=\frac{e\,\mathcal{B}}{2\,\pi\,(1-\theta\kappa)}.\label{density-I}
\end{split}
\end{align}

It is worth to say that the results for the densities in both regions may be concisely written as a unique result
 $\rho=\frac{e\,\mathcal{B}}{2\,\pi\,\lvert1-\theta\kappa\rvert}$. This density coincides with the ones obtained in \cite{Falomir:2015una} and is regular in the commutative limit.

\section{Magnetic Properties of a NC spinless fermion's gas}\label{landau-magnetization}
Let us first consider a gas consistent of spinless fermionic particles. Since a closed expression for the grand-canonical potential is not easily obtained we will consider in this section its low and high temperature limit. Our results show how the commutative results known as Landau diamagnetism and de Hass- van Alphen effect are modified by noncommutativity. Notice that the distinction between regions I and II is no longer needed, since we have already shown that the spectra and the density of states is the same.

\subsection{The high temperature limit: Landau diamagnetism}
In this section it is our aim to calculate the magnetization in the classical domain, i.e. in the high temperature limit. For that porpouse, let us now consider the grand potential density
\begin{equation}
-\beta\phi_0 = \rho\sum_{n=0}^\infty\log\left(1+ze^{-\beta\omega_c(n+\frac{1}{2})}\right).
\end{equation}
Inasmuch as $\beta\omega_c\ll1$ and average density $\bar{n}$ satisfies $\frac{\bar{n}\beta\omega_c}{\rho}\ll1$ the fugacity $z$ is also small and we may use the Taylor expansion for the logarithm\footnote{A detailed analysis of this assertion may be found in Appendix \ref{apendice-fugacidad}.}
$$\log(1+x)\sim x.$$
The resulting sum is easy to compute and we get, in terms of the dimensionless quantity $x:=\frac{\beta\omega_c}{2}$,
\begin{equation}\label{particion_landau}
-\beta\phi_0\sim \frac{z\rho}{2}\mbox{csch}(x)\,.
\end{equation}

From now on, the magnetic properties of the gas will be computed as derivatives of eq. \eqref{particion_landau} with respect to the magnetic field $B$, fixing the mean density $\bar{n}$. Before that, lets point out that --as  happens in the commutative case-- the fugacity $z$ in the high temperature limit is independent of the field strength. Indeed, we show in Appendix \ref{apendice-fugacidad} that in this regime the average density $\bar{n}$ is given by
\begin{equation}\label{density-landau}
\bar{n}= -\left.\frac{1}{ \beta }\frac{\partial (\beta\phi_0)}{\partial\mu_c}\right\rvert_{\beta,V}\sim
\frac{\rho z}{\beta \omega_c}\,.
\end{equation}

Now, we can calculate the magnetization $\mathcal{M}$ of the system as a first order derivative of the grand-canonical potential logarithm. Indeed, performing a first order expansion in the inverse temperature parameter $\beta$ and retaining only the first correction in the NC parameters $\theta$ and $\kappa$ we obtain,
\begin{equation}\label{magnetization-landau}
\begin{split}
\mathcal M:&=-\left.\frac{1}{\beta}\,\,\partial_B (\beta\phi_0)\right\rvert_{\bar{n},\beta}
\\
&\sim-\frac{\mu_B^2\bar{n}\beta}{3}\left(B + \frac{\kappa}{e} + \frac{3}{4} e B^2\theta+ O(\theta^2)\right).
\end{split}
\end{equation}
In this expression we have defined the Bohr magneton $\mu_B=\frac{e}{2\,m_e}$.

The magnetic susceptibility is analogously obtained as a second order derivative of eq. \eqref{particion_landau} with respect to the magnetic field $B$ -- its high temperature limit is
\begin{align}
\begin{split}\chi:&=\left.\frac{\partial \mathcal{M}}{\partial B}\right\rvert_{\bar{n},\beta}\\
\sim&-\frac{\mu_B^2\,\bar{n}\,\beta}{3}\left(1 + \frac{3}{2} e B \theta+O(\theta^2)\right)\,,
\end{split}\end{align}
where it behaves --as dictates Curies' Law in the usual commutative case-- as the inverse of the temperature.

Additionally, it can be straightforwardly shown that the corresponding expressions for the magnetization and the magnetic susceptibility in the 3D case coincides with the 2D ones, because the contribution from the perpendicular direction can be factorized and is $B$-independent:
\begin{align}
\begin{split}\label{magnetizacion3d-landau}
 \mathcal{M}_{3D}&\sim-\frac{\mu_B^2\bar{n}\beta}{3}\left(B + \frac{\kappa}{e} + \frac{3}{4} e B^2\theta+ O(\theta^2)\right),\\
\chi_{3D}&\sim-\frac{\mu_B^2\,\bar{n}\,\beta}{3}\left(1 + \frac{3}{2} e B \theta+O(\theta^2)\right).
\end{split}
\end{align}

We should remark that all these results are well behaved in the commutative limit $(\theta,\kappa)\rightarrow0$. Moreover, in this limit the results coincide with the magnetization and magnetic susceptibility obtained in the commutative case. An explicit computation of eq. \eqref{magnetizacion3d-landau} shows that given a field $B\sim 10^5\, \text{G}$ the NC contribution is of order $10^{-50}\text{Oe}$ for the magnetization and $10^{-56}$ for the susceptibility in gaussian units.

\subsection{The low temperature regime: the de Haas-van Alphen effect}

In 1930 de Haas and van Alphen were the first to observe a quasi periodic variation of $\chi$ with the strength of the applied field $B$ for bismouth's crystals at low temperatures\cite{deHaas}. At $T=0$ temperature this effect may be qualitatively understood by following Peierls \cite{Peierls} -- recall that from the definition of the density $\bar{n}$ we have
\begin{align}\label{density-fugacity-vanalphen}
\bar{n}&=\rho\,\sum_{n=0}^{\infty} \frac{z\,e^{-\beta\omega_c\left(n+\frac{1}{2}\right)}}{1+z\,e^{-\beta\omega_c\left(n+\frac{1}{2}\right)}},
\end{align}
and interpret each term in this sum as a level population.

Then, define $\mathcal{B}_0:=\frac{\bar{n}}{\rho}\,\mathcal{B}$ and consider a temperature slightly higher than zero and a field strength $\mathcal{B}$ which is bigger than $\mathcal{B}_0$. Then from eq. \eqref{density-fugacity-vanalphen} it can be seen that $z\sim \frac{\mathcal{B}_0}{\mathcal{B}}\,e^{\frac{\beta\omega_c}{2}}$, such that in the strict $T=0$ case the only populated level is the ground level. If the field strength $\mathcal{B}$ is decreased, from eqs. \eqref{density-II}-\eqref{density-I} it follows that $\rho$ decreases and the ground state has not enough available places to contain all the fermions. As a result, the $n=1$ state --i.e. the first excited state-- begins to be populated and therefore the fugacity should behave at least as $z\sim e^{\frac{3\beta\omega_c}{2}}$. Of course higher levels will begin to be populated as $\mathcal{B}$ keeps decreasing.

\begin{figure}[h]\label{graph-vanalphen} 
\begin{center}
\begin{minipage}{0.45\textwidth}
\includegraphics[width=\textwidth]{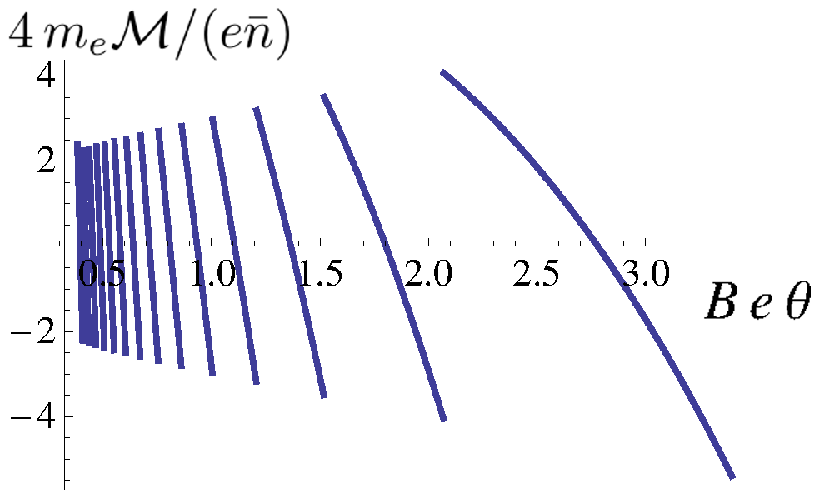} 
\end{minipage}
\begin{minipage}{0.45\textwidth}
\includegraphics[width=\textwidth]{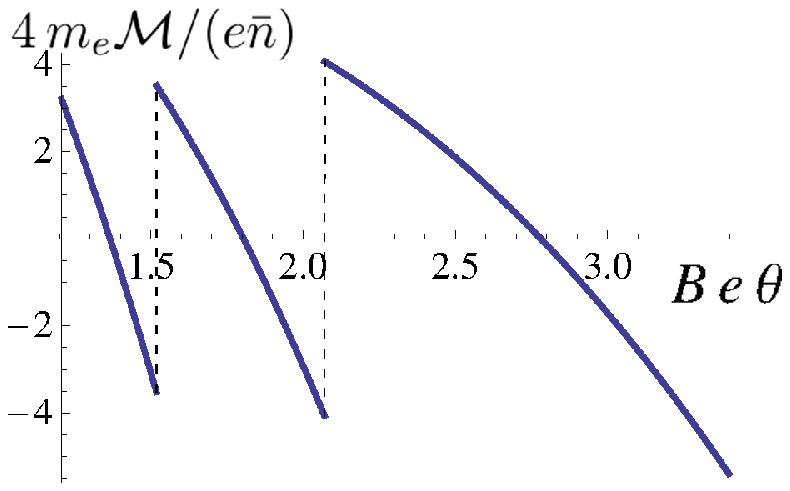} 
\end{minipage}

\caption{Plot of the reduced magnetization $4\,m_e\mathcal{M}/(e\bar{n})$ vs $e\theta B$. The graphic on the right is a zoom of the graphic on the left. }
\end{center}
\end{figure}

The consequence of this reasoning is that the partition function, the magnetization and the susceptibility become piecewise functions of the strength field $\mathcal{B}$ -- the latter two can be casted in the form
\begin{equation}\label{eqn-vanalphen}
\begin{array}{l}
 \displaystyle
\mathcal{M}= -\frac{e\bar{n}}{4\,m_e}\left(2+e\,\theta\,B\right) \left(2\frac{\mathcal{B}}{\mathcal{B}_0}j(j+1) -(2j+1)\right)\,,
\\ \\ \displaystyle
\mathcal{\chi}=  -\frac{e\bar{n}}{4\,m_e\mathcal{B}_0}\left[\left(4+2e\theta(B+\mathcal{B})+e^2\theta^2\,B^2\right)j(j+1) -e\theta\mathcal{B}_0(2j+1) \right],
\end{array}
\end{equation}
for $\frac{1}{j+1}<\frac{\mathcal{B}}{\mathcal{B}_0}<\frac{1}{j}$ where $\,j\in\mathbb{N}_{0}$. Once more the commutative limit is regular. Notice that for null magnetic field there is a constant $\kappa-$dependent remanent magnetization of the system. This would be a measurable quantity in a model where its value were substance dependent.
\\
In Figure \ref{graph-vanalphen} we have plotted eqn. \ref{eqn-vanalphen} for the reduced magnetization $4\,m_e\mathcal{M}/(e\bar{n})$ vs $e\theta B$, setting $\kappa=0$ and $\theta \bar{n}=1$. As may be seen from the zoom, the effects of noncommutativity becomes increasingly important as j decreases.

\section{Magnetic properties of a NC spin-$1/2$ fermions' gas}\label{section.pauli}
We now turn our attention to the case where fermions have spin $1/2$ and its Hamiltonian includes an interaction term $\pm\mu_0\,B_{spin}$ between the spin of the particles and the external magnetic field $B_{spin}$. We shall then follow one of two paths: we may either consider that our model comes from the generalization of a Schr\"odinger operator by replacing the usual product by the Moyal product, or propose that it is nothing but an effective model of a NC quantum field theory. The difference is that in the first case the interaction is through the magnetic field $B$, while in the second one we should consider the effective magnetic field $\mathcal{B}$, cf. eq. \eqref{effective-magnetic-field}, computed using the Seiberg-Witten' map.

In either case, we need to compute the grand potential density
\begin{align}
-\beta\phi=\rho\sum_{s=\pm 1}\sum_{n=0}^{\infty} \log(1+e^{-\beta \omega_c n+\alpha_s}),
\end{align}
where we have included in the $\alpha_s$ parameter the interaction of the spin with a field $B_{spin}$ that will be later appropriately chosen:
\begin{align}\label{alpha_s}
 \alpha_s=\beta\mu_c+\beta\mu_0 s {B}_{spin}-\frac{\beta\omega_c}{2}.
\end{align}

\subsection{The high temperature regime}
In the high temperature regime we may employ an expansion of the logarithm similar to the one used in section \ref{landau-magnetization}. Indeed, for high temperatures and low densities, i.e. $\beta(\omega_c+\mu_0\,B_{spin})\ll 1$ and $\frac{\bar{n}\beta}{\rho}(\omega_c+\mu_0\,B_{spin}) \ll1$, we obtain
\begin{align}
\begin{split} -\beta \phi_0 &\sim\rho\sum_{s=\pm 1}\sum_{n=0}^{\infty} e^{-\beta \omega_c n+\alpha_s}\\
&= -2\beta \cosh(\beta\mu_0\mathcal{B}) \phi_0,
\end{split}
\end{align}
where we have defined $\phi_0$ as the logarithm of the partition function in absence of the spin-field interaction. Of course the magnetization may be still computed in terms of the partition function, leading to
\begin{align}\label{magnetization-pauli-t-high}
 \mathcal{M}&\sim -2\phi_0\frac{\partial \cosh(\beta\mu_0\mathcal{B})}{\partial B} -2\frac{\partial \phi_0}{\partial B} \cosh(\beta\mu_0\mathcal{B}) .
\end{align}

The RHS of eq. \eqref{magnetization-pauli-t-high} tells us that the magnetization in the high temperature limit consists of two terms -- the first  one corresponds to the Pauli paramagnetism while the second one is nothing but the Landau diamagnetism we have already studied in section\footnote{Actually there is a factor two of difference which arises in eq. \eqref{magnetization-pauli-t-high} because of the two possible spin states.} \ref{landau-magnetization}. The validity of this assertions rests in the first order approximation of $\beta$ studied in Appendix \ref{apendice-fugacidad} where $z$ is $B$-independent. From the other side, in a higher order expansion of $\beta$, there will be corrections due to the $B$ dependence in $z$. The expressions for $\mathcal{M}^P$ and $\chi^{P}$, the Pauli paramagnetism contributions to the magnetization and
magnetic susceptibility, finally are
\begin{align}
 \begin{split}\label{magnetizacion3d.pauli}
\mathcal{M}^P=\mathcal{M}^P_{3D}
&\sim \mu_0^2\bar{n}B\beta\left(1-\frac{3}{4}Be\theta\right)+\frac{2\mu_0^2\bar{n}}{e}\kappa\beta,\\
\chi^{P}=\chi^{P}_{3D}&\sim \mu_0^2\bar{n}\beta\left(1-\frac{3}{2}Be\theta\right).
\end{split}
\end{align}
As in the case of Landau's diamagnetism, the expressions for the 3D problem coincide with their 2D counterpart.
These expressions show that Pauli's paramagnetism contributions is three times bigger than Landau's diamagnetism contribution. It is also important to notice from \eqref{magnetizacion3d.pauli} that NC corrections are diamagnetic and of the same order of magnitude as those obtained for Landau's diamagnetism.

\subsection{The low temperature regime}
In the low temperature regime --i.e. when the Fermi energy of the system $\epsilon_F$ is such that $\beta\epsilon_F\gg1$ -- the partition function may be computed by using Abel-Plana formula to rewrite the sum into integrals. A detailed calculation may be found in Appendix \ref{apendice-abel-plana} and yields
\begin{align}
-\beta\phi_0= \rho \sum_{s=\pm} \left(\frac{\beta\omega_c}{12}+\frac{3\alpha_s^2+\pi^2}{6\beta\omega_c}+\text{coth}\left(\frac{\beta\omega_c}{2}\right) e^{-\alpha_s}+O(e^{-2\alpha_s})\right).
\end{align}

As we have already done before, we shall obtain an expression for the chemical potential $\mu_c$ in terms of the mean density $\bar{n}$ -- this will aid us to compute the magnetization and the magnetic susceptibility as derivatives of the partition function. Therefore we write
\begin{align}\label{n-medio-pauli-low}
 \bar{n}= \rho\sum_{s=\pm} \frac{2\mu_c+2s\mu_0B_{spin}-\omega_c}{2\omega_c}+\text{coth}\left(\frac{\beta\omega_c}{2}\right) e^{-\beta(\mu_c+s\mu_0B_{spin}-\frac{\omega_c}{2})}+O(e^{-2\alpha_s}).
\end{align}
In the strict $T=0$ limit, it is clear from \eqref{n-medio-pauli-low} that
\begin{align}
 \mu_c(T=0)=\left(\frac{\bar{n}}{\rho}+1\right)\frac{ \omega_c}{2}.
\end{align}
One may also observe from eq. \eqref{n-medio-pauli-low} that, as the temperature increases from zero,  the states with a spin component parallel to the direction of the field are populated faster than the other ones. If we now consider a small variation in temperature, assuming $\mu_c$ is a smooth function of temperature we may use eq. \eqref{n-medio-pauli-low} once more to obtain
\begin{align}
 \mu_c(T)=\mu_c(T=0)-\frac{2}{\rho} \text{coth}\left(\frac{\beta\omega_c}{2}\right) \cosh(\beta\mu_0\omega_c)e^{-\frac{\beta\omega_c}{2}\left(\frac{\bar{n}}{\rho} -1\right)}+\cdots.
\end{align}
We thus see that energy states for the system other than $\mu(T=0)$ are highly supressed in the low temperature limit.

Finally, it may be shown using the definition \eqref{magnetization-landau} as the derivative of the grand partition density w.r.t. the magnetic field $B$ that the magnetization has the following expression for $T=0$:
\begin{align}\label{magnetization-pauli-T0}
 \begin{split}\mathcal{M}&=\frac{e^2}{6\pi\,m_e\lvert1-\theta\kappa\rvert}\mathcal{B}\partial_B\mathcal{B}+\frac{2\rho\mu_0^2}{\omega_c} {B}_{spin}\frac{\partial {B}_{spin}}{\partial B}\\
&=\frac{e^2}{6\pi\,m_e\lvert1-\theta\kappa\rvert}\mathcal{B}\left(1+\frac{e\theta B}{2}\right)+\frac{\mu_0^2\,m_e}{\pi\lvert1-\theta\kappa\rvert}
\begin{cases}
\left(1+\frac{\theta e B}{2}\right)\mathcal{B}, &\text{ for } B_{spin}=\mathcal{B}\\
B, &\text{ for } B_{spin}=B
\end{cases}.
\end{split}
\end{align}
From this result it is clear that the NC Landau problem has two independent corrections when it is compared with the commutative case: one coming from the density of states and the other one introduced by the Seiberg-Witten map.

\section{Discussion and Conclusions}
In this paper we have studied the thermodynamical behaviour of a fermion gas subject to an external constant magnetic field in 2D and 3D NC space. Indeed, we have obtained the exact expressions in terms of the NC $\theta$-$\kappa$ parameters for the magnetization and the magnetic susceptibility for the diamagnetic and paramagnetic contribution.

These expressions turn out to be regular in $\theta$ so the analysis reduces to considering their first order contributions. Notably, they have the same sign and reinforce each other. An explicit computation shows that, given an electronic gas subject to a magnetic field of $10^5\,\text{G}$, a typical density of $\bar{n}\approx10^29\text{ m}^{-3}$ for electrons in metals and a NC parameter of order\footnote{Planck's length is defined as $\ell_P=\sqrt{\hbar G/c^3}$, where $\hbar$ is Planck's reduced constant, $G$ is the gravitational constant and $c$ is the speed of light in vacuum.} $\theta\sim\ell_P^2\sim (10^{28}\, \text{eV})^{-2}$, this contribution is of order $10^{-50}\,\frac{\text{emu}}{\text{cm}^3}$ for the magnetization and $10^{-56}$ for the magnetic susceptibility. Moreover, the magnetic field itself would be corrected by an amount of $10^{-49}\,\text{G}$. Should these be the NC contributions they will be extremely hard to measure -- e.g. these contributions would be equivalent to the addition of 
the moment of a free electron to a sample of $10^{29}\,\text{cm}^3$ of a metal.

Although in the last years there has been a great improvement in experimental tecniques, these figures are beyond the current maximum attainable resolution of $10^{-13}\,\text{G}$  for a SQUID magnetometer (\cite{Mester:2000nfc}). Considering this resolution we may obtain an upper bound for the NC parameter $\theta\lesssim (10\,\text{Gev})^{-2}$. Notice that this bound is obtained from the behaviour of a macroscopic body while bounds are usually obtained either from particle's experiments or astronomical data. Our result is weak as the former -- e.g., in \cite{Chaichian:2000si} the bound $\theta\lesssim 10^{-6}\lambda_e^2\alpha\sim (1\,\text{GeV})^{-2}$ is obtained\footnote{$\lambda_e\sim 2\,10^{-12}\, \text{m}$ is the electron Compton wavelength and $\alpha\sim \frac{1}{137}$ is the fine-structure constant.}.

Probably one of the best options to study the feasability of these predicted NC effects are magnetars -- neutron stars whose magnetic fields are expected to be as big as $10^{14}\text{G}$ and emit high energy electromagnetic radiation through GRB \cite{Castorina:2004rc}. This analysis is work in progress.

\section{Acknowledgments}
The authors acknowledges financial support from the Physics Department of the UNLP and from the UNLP through the grant 11/X615.
\\
They also want to express their gratitude to Falomir H. for making important comments that allowed them to improve the present work.

\begin{appendix}

\section{The fugacity in the $\beta\sim0$ regime}\label{apendice-fugacidad}
In the high temperature regime --i.e. in the inverse temperature $\beta\sim0$ regime-- we may obtain the asymptotic expansion of the fugacity $z$ in terms of $\beta$ and the density of particles $\bar{n}$. As a first step, we compute $\bar{n}$ as the derivative of the partition function with respect to $\mu_c$ as $\beta$ remains constant:
\begin{align}\label{density-fugacity}
\bar{n}
&= \rho\,z\sum_{n=0}^{\infty} \frac{e^{-\beta\omega_c\left(n+\frac{1}{2}\right)}}{1+z\,e^{-\beta\omega_c\left(n+\frac{1}{2}\right)}}.
\end{align}

From this expression it can be seen that as the temperature becomes larger $z$ must tend to zero in order for $\bar{n}$ to remain finite. Using this fact, one may expand the denominator\footnote{Note that in order to keep $\bar{n}$ finite  the relation $z\exp^{-\beta\omega_c(n+1/2)}<1$ must hold.} in \eqref{density-fugacity} and then recast the density of particles by performing a resummation:
 \begin{align}\label{density-fugacity-recast}
\bar{n}&=- \rho\,\sum_{r=1}^{\infty}\ \frac{\left(-ze^{-\beta \omega_c/2}\right)^{r}}{1-{e^{-\beta \omega_c r}}}.
\end{align}

Now, it is known that every series like \eqref{density-fugacity-recast} may be inverted by using the Lagrange inversion theorem. The final result is
\begin{align}\label{fugacity-density}
 z=-e^{\beta \omega_c/2}\sum_{n=1}^{\infty} \frac{1}{n!} \left( \sum_{k=1}^{n-1}\frac{(-1)^{k} (n+k-1)!}{(n-1)!} B_{n-1,k}(\hat{f}_1,\cdots,\hat{f}_{n-k}) \right) \left(-\frac{f\,\bar{n}}{\rho} \right)^n,
\end{align}
where we have made use of the Bell polynomials $B_{n,k}$, and defined the following functions of $\beta$:
\begin{align}
\begin{split}\hat{f}_k:&=\frac{1-e^{-\beta\omega_c}}{(k+1)(1-e^{-\beta\omega_c(k+1)})},\\
f:&= 1-e^{-\beta\omega_c}.
\end{split}
\end{align}

It should be noticed that expression \eqref{fugacity-density} is well-behaved in the limit $\beta\rightarrow0$ and provides us with the following expansion for the fugacity $z$:
\begin{align}
z
\sim \frac{\bar{n}\beta\omega_c}{\rho} +O(\beta^2).
\end{align}

\section{On using Abel-Plana formula to compute Pauli's paramagnetism in the low temperature regime}\label{apendice-abel-plana}
In this appendix we will show how to obtain an analytical result for the partition function considered in section \ref{section.pauli} for the Pauli paramagnetism' problem,
\begin{align}
-\beta\phi=\rho\sum_{s=\pm 1}\sum_{n=0}^{\infty} \log(1+e^{-\beta \omega_c n+\alpha_s}),
\end{align}
where the parameter $\alpha_s$ has been defined in eq. \eqref{alpha_s}.

For a while we will forget about the index $s$ which labels the spin up and down components -- if we just focus on the sum over the $n$ index we can write according to Abel-Plana Formula
\begin{multline}\label{abel-plana}-\beta\phi= \int_0^{\infty}dx  \log (1+e^{-\beta\omega_c x+\alpha_s}) + \frac{1}{2} \log(1+e^ {\alpha_s})\\+ i\int^{\infty}_0 \frac{1}{e^{2\pi x}-1} \log\left(\frac{1+e^{-i\beta\omega_c x+\alpha_s}}{1+e^{i\beta\omega_c x+\alpha_s}}\right).
\end{multline}

The first integral on the RHS of \eqref{abel-plana} may be splitted in the following two, which may be easily computed for $\beta\,\epsilon_F\gg 1$:
\begin{align}\label{abel-plana-primera}
 \begin{split}\int_0^1 dx \frac{\alpha}{\beta\omega_c} \log(1+e^{(1-x)\alpha})&=\frac{6\alpha^2+\pi^2+12Li_2(e^{-\alpha})}{12\beta\omega_c},\\
 \int_0^{\infty}dx \log(1+e^{-\beta\omega_c x})&=\frac{\pi^2}{12\beta\omega_c},
\end{split}
\end{align}
where $Li_2(\cdot)$ is the polylogarithm function of order 2.

On the other hand, the second integral in the RHS of \eqref{abel-plana} may be expanded perturbatively in the low temperature regime, viz. when $\beta\epsilon_F\gg 1$ or equivalently $\alpha_s\gg1$. In effect, the argument of the logarithm can be recasted as
\begin{align}
 \frac{1+e^{-i\beta \omega_cx+\alpha}}{1+e^{i\beta \omega_cx+\alpha}} =e^{-2ix} \frac{1+e^{i\beta\omega_c x-\alpha}}{1+e^{-i\beta\omega_c x-\alpha}}.
\end{align}
Thus, since the $x$-dependent part is oscillatory the Taylor expansion for $\log(1+x)$ around $x=0$ may be used. The result is
\begin{align}\label{abel-plana-segunda}
 i\int^{\infty}_0 \frac{1}{e^{2\pi x}-1} \log\left(\frac{1+e^{-i\beta\omega_c x+\alpha}}{1+e^{i\beta\omega_c x+\alpha}}\right) =\frac{\beta\omega_c}{12}+\frac{1-\beta\omega_c\text{coth}(\beta\omega_c/2)}{\beta\omega_c}e^{-\alpha}+O(e^{-2\alpha})
\end{align}
Adding expressions \eqref{abel-plana}, \eqref{abel-plana-primera} and \eqref{abel-plana-segunda}, and reintroducing the spin' index s, we finally obtain for the partition function in the low temperature regime
\begin{align}
-\beta\phi= \sum_{s=\pm} \left(\frac{\beta\omega_c}{12}+\frac{3\alpha_s^2+\pi^2}{6\beta\omega_c}+\text{coth}\left(\frac{\beta\omega_c}{2}\right) e^{-\alpha_s}+O(e^{-2\alpha_s})\right).
\end{align}

\end{appendix}

\vspace{2cm}


\end{document}